\documentclass[lettersize,journal]{IEEEtran}
\usepackage{times}
\usepackage{amsmath,amsfonts,gensymb, algorithmic, array}
\usepackage[caption=false,font=normalsize,labelfont=sf,textfont=sf]{subfig}
\usepackage{textcomp, stfloats, url,verbatim}
\def\BibTeX{{\rm B\kern-.05em{\sc i\kern-.025em b}\kern-.08em
    T\kern-.1667em\lower.7ex\hbox{E}\kern-.125emX}}
\usepackage{balance,graphicx,multirow, multicol, amssymb, amsmath,changepage, xspace, booktabs,pifont,subfig} 
\usepackage{hyperref}

\begin{document}

\title{CST-former: Multidimensional Attention-based Transformer for Sound Event Localization and Detection in Real Scenes}

\author{Yusun Shul, \IEEEmembership{Student Member, IEEE}, Dayun Choi, \IEEEmembership{Student Member, IEEE}, and Jung-Woo Choi, \IEEEmembership{Member, IEEE}
\thanks{This work was supported by the National Research Foundation of Korea (NRF) grant funded by the Ministry of Science and ICT of Korea government (MSIT) (No. RS-2024-00337945), and the BK21 FOUR program through the NRF funded by the Ministry of Education of Korea.}
\thanks{Yusun Shul, Dayun Choi, and Jung-Woo Choi are with the Korea Advanced Institute of Science and Technology (KAIST), Daejeon 34141, South Korea (e-mail:
shulys@kaist.ac.kr; cdy3773@kaist.ac.kr; jwoo@kaist.ac.kr).}}

\markboth{
}
{Shell \MakeLowercase{\textit{et al.}}: Bare Demo of IEEEtran.cls for IEEE Journals}
\maketitle

\begin{abstract}
Sound event localization and detection (SELD) is a task for the classification of sound events and the identification of direction of arrival (DoA) utilizing multichannel acoustic signals. 
For effective classification and localization, a channel-spectro-temporal transformer (CST-former) was suggested. CST-former employs multidimensional attention mechanisms across the spatial, spectral, and temporal domains to enlarge the model's capacity to learn the domain information essential for event detection and DoA estimation over time. In this work, we present an enhanced version of CST-former\footnotemark[1] with multiscale unfolded local embedding (MSULE) developed to capture and aggregate domain information over multiple time-frequency scales. Also, we propose finetuning and post-processing techniques beneficial for conducting the SELD task over limited training datasets. In-depth ablation studies of the proposed architecture and detailed analysis on the proposed modules are carried out to validate the efficacy of multidimensional attentions on the SELD task. Empirical validation through experimentation on STARSS22 and STARSS23 datasets demonstrates the remarkable performance of CST-former and post-processing techniques without using external data.
 

\end{abstract}


\footnotetext[1]{Previous version of this paper is available in \cite{CST_former}.}

\IEEEpeerreviewmaketitle

\section{Introduction}

\IEEEPARstart{S}{ound} event localization and detection (SELD) refers to the multi-task for sound event detection (SED) and direction of arrival estimation (DoAE) in a three-dimensional (3-D) space over time~\cite{2018DCASE_baseline, 2019overview_TASLP}. 
The Detection and Classification of Acoustic Scenes and Events (DCASE) challenge task 3~\cite{2018DCASE_baseline} aims to develop a deep neural network (DNN)-based model for SELD. 
To accomplish SED and DoAE tasks simultaneously, several multi-task models, e.g., a network producing multiple losses~\cite{2019overview_TASLP, 2020Du_1st, 2020Nguyen_2nd, 2020Park_5th}, have been proposed. 

As the output format for the joint estimation of SED and DoAE, Shimada et al. introduced the activity-coupled cartesian DoA (ACCDOA)~\cite{accdoa} and multi-ACCDOA~\cite{multiaccdoa} that couple SED and DoAE outputs as the length and direction of a vector. Due to their benefits in handling complex sound scenes, ACCDOA and multi-ACCDOA
have been popularly employed
~\cite{seresnet_gru,ResnetConformer, 2022DCASE_baseline}.

Until 2021, the primary focus of the DCASE challenge and suggested models~\cite{2020Du_1st, 2020Nguyen_2nd, einv2} had centered on developing SELD algorithms for synthetic datasets. 
To develop practical SELD models applicable to real-world scenarios, the DCASE challenge task 3 has provided real recordings since 2022, specifically STARSS22~\cite{2022DCASE_baseline} and STARSS23~\cite{data_starss23}, recorded in various room environments. Additionally, synthetic data (\emph{synth-set})~\cite{data_DCASE2022synth} generated by convolving single channel data FSD50K~\cite{FSD50K} and TAU-SRIRs~\cite{data_tau_srirDB} have been provided to address the lack of sufficient real audio recordings. Nevertheless, achieving robust SELD performance through training over a limited dataset remains challenging due to unseen DoAs, unseen reverberations, and various signal-to-noise ratios (SNRs) of the data recorded in real-world environments. 

To address the challenges of SELD in real scenes, DNN architectures have evolved to encode information more effectively. 
Models like convolutional recurrent neural network (CRNN) of the DCASE baseline~\cite{2022DCASE_baseline}, ResNet-Conformer~\cite{ResnetConformer} and event independent network V2 (EINV2)~\cite{einv2, einv2_2022} 
employ temporal attention mechanisms or gated recurrent units (GRUs) to capture temporal context (Table~\ref{tab: Models_table}). 
However, these approaches primarily rely on spectral and channel embeddings as input to the temporal attention mechanisms, which may hamper extracting meaningful relations across spectral or spatial domains. 
While data augmentation and the incorporation of diverse external datasets have been shown to significantly enhance SELD performance ~\cite{ResnetConformer, ACS_TASLP}, these approaches are resource-intensive, and securing external data may not always be feasible. Consequently, there is a critical need to design DNN architectures capable of achieving robust performance under a constrained set of training data.

Research in other fields, such as speech enhancement~\cite{TPARN, Dasformer, CMGAN}, speech separation~\cite{TFgridnet, sepformer}, and video classification~\cite{video_dividedatten}, has demonstrated the benefits of applying independent attention or RNN mechanisms across different domains or dimensions with various methods. 
These mechanisms use separate attention layers employed for each domain by permuting the data dimensions~\cite{TPARN, Dasformer, CMGAN, video_dividedatten}. 
In addition, Wang et al.~\cite{TFgridnet} introduced an unfolding mechanism to reduce the sequence dimension and computation cost accompanied by separated attention layers. Subakan et al.~\cite{sepformer} considered chunking of input waveforms to realize intra- and inter-transformers modeling short- and long-term dependencies. 
These domain-dependent attention mechanisms hold potential for application in SELD to extract valuable information from multidimensional data. 

To this end, we previously introduced the divided spectro-temporal (DST) attention~\cite{ST_attention, ST_former}, which employs separated attention mechanisms for temporal and spectral domains. While DST attention-based models achieved significant improvement over the baseline of the DCASE 2023 challenge, they exhibited a notable limitation: channel information was utilized solely as an embedding of spectral and temporal sequences.
Building on this, the recently proposed CST-former~\cite{CST_former} applies distinct multidimensional attention across three domains --- time, frequency, and channel --- by leveraging an unfolded local embedding (ULE) operation to extract channel attention (CA) embeddings efficiently. 
However, this preliminary study does not explain how this architecture successfully works for the SELD task. Moreover, the ULE technique extracting CA embeddings from local T-F grids or fixed-size patches may be limited in capturing features across diverse T-F sizes. Compared to other works, the preliminary work also lacks key refinement techniques to improve the SELD performance, such as finetuning event detection threshold and applying post-processing strategies during inference.

In this follow-up research, therefore, we validate the effectiveness of the CST-former through in-depth ablation studies and analysis. Furthermore, we introduce advanced estimation techniques to further enhance performance. These techniques include multiscale ULE (MSULE), vector threshold masking (VTM), inference overlapping (IO), and clustered-track augmented inference (CTAI).

The contributions of this paper are summarized as follows:
\begin{itemize} 
\item We present a detailed architecture of the CST-former regarding the multidimensional features of SELD: spatial, spectral, and temporal characteristics. 
Embedding generation method, ULE, is presented in detail, along with the novel multiscale approach (MSULE).
\item We propose 
finetuning method (VTM) for decreasing false negatives (FNs) in the event detection, and post-processing technique (CTAI) for resolving the track permutation issue in the test-time augmentation (TTA).
\item We also demonstrate the effectiveness of the proposed CST-former on the SELD task through more in-depth ablation studies and analyses. Especially, the benefits of multidimensional attention on SELD are thoroughly inspected and verified with analyses per class, per room, and on channel attention maps.
\end{itemize}

The paper is organized as follows. Section~\ref{sec:related} briefly presents the related works. In Section~\ref{sec:proposed}, we introduce the details of the proposed transformer architecture and the embedding generation method for CA, ULE, along with MSULE. Also, the VTM and inference time processing (IP) techniques are proposed to enhance the SELD performance of the CST-former further. Then, the experimental results are demonstrated in Section~\ref{sec:results}, containing parameter studies and comparison with state-of-the-art models. Analyses are demonstrated in Section~\ref{sec:discussion} to verify the effectiveness of the proposed architecture, and Section~\ref{sec:conclusion} concludes the paper.

\begin{table}[t]
\caption{Comparison of model architectures. Front, middle, and end pooling locations indicate pooling after the convolution layers in the encoder, right before the decoder, and right before the fully connected layer at last, respectively.}
\label{tab: Models_table}
\resizebox{\columnwidth}{!}{%
\setlength\tabcolsep{2.5pt}
\renewcommand{\arraystretch}{1.2}
\begin{tabular}{lcccccc} 
\hline
Model&Encoder&Decoder&\renewcommand{\arraystretch}{1.0} \begin{tabular}[c]{@{}c@{}} Attention\\Domain\end{tabular}&\renewcommand{\arraystretch}{1.0} \begin{tabular}[c]{@{}c@{}}Pooling\\Location\end{tabular}&Output Type & \renewcommand{\arraystretch}{1.0} \begin{tabular}[c]{@{}c@{}} Parameter\\Size\end{tabular}\\ \hline \hline
\begin{tabular}[l]{@{}l@{}}2022 Baseline\\ (CRNN)~\cite{2022DCASE_baseline}\end{tabular} &CNN&GRU&-&Front&\renewcommand{\arraystretch}{1.0} \begin{tabular}[c]{@{}c@{}} Multi-\\ACCDOA\end{tabular} & 0.60M \\  \hline
\begin{tabular}[l]{@{}l@{}}2023 baseline\\ (CRNN)~\cite{dcase2023_baseline}\end{tabular} &CNN&\renewcommand{\arraystretch}{1.0} \begin{tabular}[c]{@{}c@{}} GRU,\\MHSA\end{tabular}&Time&Front&\renewcommand{\arraystretch}{1.0} \begin{tabular}[c]{@{}c@{}} Multi-\\ACCDOA\end{tabular}&0.74M\\ \hline
\begin{tabular}[l]{@{}l@{}}ResNet-\\ Conformer~\cite{ResnetConformer}\end{tabular} &ResNet&Conformer&Time&\begin{tabular}[c]{@{}c@{}}Front\\Middle\\End\end{tabular}&\renewcommand{\arraystretch}{1.0} \begin{tabular}[c]{@{}c@{}} Multi-\\ACCDOA\end{tabular}&58M\\ \hline
EINV2~\cite{einv2}&CNN & Conformer&Time&Front&Multi-task&85M\\ \hline
DST Attention~\cite{ST_attention}& CNN& \renewcommand{\arraystretch}{1.0} \begin{tabular}[c]{@{}c@{}} DST-\\MHSA\end{tabular}& \renewcommand{\arraystretch}{1.0} \begin{tabular}[c]{@{}c@{}}Frequency\\Time\end{tabular} & Front & \renewcommand{\arraystretch}{1.0} \begin{tabular}[c]{@{}c@{}} Multi-\\ACCDOA\end{tabular}&0.30M \\ \hline
\begin{tabular}[l]{@{}l@{}}CST-former\end{tabular} &CNN&\begin{tabular}[c]{@{}c@{}}CST-\\ transformer\end{tabular}&\renewcommand{\arraystretch}{1.0} \begin{tabular}[c]{@{}c@{}c@{}}Channel \\Frequency\\Time\end{tabular}&\begin{tabular}[c]{@{}c@{}}Front\\Middle\\End\end{tabular} & \renewcommand{\arraystretch}{1.0} \begin{tabular}[c]{@{}c@{}} Multi-\\ACCDOA\end{tabular} &  \begin{tabular}[c]{@{}c@{}}0.39M\\ 1.88M\end{tabular}\\ \hline 
\end{tabular}
}
\end{table}

\section{Related Works} \label{sec:related}
First of all, the common structure most of the previous models for SELD~\cite{ResnetConformer,einv2,dcase2023_baseline} share is temporal attention. Specifically, the baseline model for 2023 DCASE challenge task 3 (2023 baseline)~\cite{dcase2023_baseline} incorporates several key components: convolutional blocks, bi-directional gated recurrent units (GRU), and multi-head self-attentions (MHSA). The convolutional blocks are responsible for expanding the number of input channels and employing max pooling after each convolutional operation to reduce the spectral dimension.
Temporal pooling is also employed to align the time dimension of the target frame. In the case of ResNet-Conformer~\cite{ResnetConformer}, ResNet18 is utilized for input feature encoding, then, a conformer structure for extracting temporal features that employ spectral and spatial information as the embedding for the temporal sequence. Moreover, EINV2~\cite{einv2} encodes the features for SED and DoAE separately with two parallel convolutional encoders, then uses track-wise conformers, each of which extracts temporal features for separate track output. The track-wise conformers of EINV2 also employ spectral and spatial information as embedding of temporal sequence.

\subsection{Input}
The input data is the $M$-channel data consisting of log-mel spectrograms of four-channel First-order-Ambisonics (FoA) signals and three intensity vectors (IVs)~\cite{IV_etri, IV_NTT} with $T$ time frames and $F$ frequency bins. The input feature is presented as
\begin{equation} \label{Eq:input}
\mathbf{X}\!\in\!{\mathbb{R}^{M\!\times\!T\!\times\!F}}.
\end{equation}

\subsection{Encoder}
Although the specific structures for the encoder of three previous models vary in detail, e.g., convolutional encoder, Resnet encoder, and parallel convolutional encoder, all the encoders amplify the number of input channels from $M$ to $C$ and encode local T-F relations. The encoded output from any encoder structure is presented as
\begin{equation} \label{Eq:EncOut}
\mathbf{z}
=E(\mathbf{X})\in{\mathbb{R}^{C\times T^\prime \times F^\prime}}
\end{equation} where $C$ is the number of filters of the encoder $E$, and $T^\prime$ and $F^\prime$ are encoded time and frequency dimensions determined by pooling after convolution in $E$.

All the three previous models reshape the encoded output $\mathbf{z}$ 
to new embedding for the $l$-th attention layer as
\begin{equation} \label{Eq:TAttenInput}
\mathbf{e}^{(l)}_{t} \in \mathbb{R}^{T^\prime \times D}
\end{equation}
where the embedding dimension is $D=(CF^\prime)$ and the sequence is temporal axis. Therefore, the joint dimension of the encoded channel and frequency, $D$, is used as the embedding of the temporal sequence. 

\subsection{Decoder}
The previous models have varying decoders such as bi-directional GRU followed by MHSA, and Conformer. Both structures contain MHSA and the features from MHSA are calculated as follows. 
\subsubsection{Temporal attention}
With $L$ number of attention layers and $H$ number of attention heads, query/key/value of layer $l$ with the head $h$ for temporal attention (T-Attention) are computed as
\begin{equation} \label{Eq:tquery}
\mathbf{q}^{(l,h)}_{t} = \mathbf{e}^{(l)}_{t}\mathbf{W}_{Q}^{(l,h)} \in \mathbb{R}^{T^\prime \times D_h}, 
\end{equation}
\begin{equation} \label{Eq:tkey}
\mathbf{k}^{(l,h)}_{t} = \mathbf{e}^{(l)}_{t}\mathbf{W}_{K}^{(l,h)} \in \mathbb{R}^{T^\prime \times D_h}, 
\end{equation}
\begin{equation} \label{Eq:tvalue}
\mathbf{v}^{(l,h)}_{t} = \mathbf{e}^{(l)}_{t}\mathbf{W}_{V}^{(l,h)} \in \mathbb{R}^{T^\prime \times D_h}, 
\end{equation} where the dimension of each head attention index $D_h$ is $D/H$, and $\mathbf{W}_{Q}^{(l,h)},\mathbf{W}_{K}^{(l,h)},\mathbf{W}_{V}^{(l,h)}\in\mathbb{R}^{D\times D_h}$ are linear operators.

From the query of Eq.~\ref{Eq:tquery} and key of Eq.~\ref{Eq:tkey}, the temporal attention is calculated as
\begin{equation} \label{Eq:TempAtten}
\mathbf{A}^{(l,h)}_{t}=\mathrm{Softmax} \left(\frac{\mathbf{q}^{(l,h)}_{t} \times {\mathbf{k}^{(l,h)}_{t}}^\intercal}{\sqrt{D_h}} \right) \in \mathbb{R}^{T^\prime \times T^\prime}. 
\end{equation} 
The attention of head $h$ and layer $l$ is multiplied by the value of Eq.~\ref{Eq:tvalue}, 
\begin{equation} \label{Eq:TempSAtten}
\mathbf{SA}^{(l,h)}_{t}=\mathbf{A}^{(l,h)}_{t}\times \mathbf{v}^{(l,h)}_{t} \in \mathbb{R}^{T^\prime\times D_h}.
\end{equation}
Then, $\mathbf{SA}^{(l,h)}_{t}$ of $H$ heads are concatenated, then multiplied with linear operator $\mathbf{W}^{(l)}\in\mathbb{R}^{D\times D}$, resulting in T-MHSA of block $l$ as
\begin{equation} \label{Eq:TMHSA}
\mathbf{MSA}^{(l)}_{t}=[\mathbf{SA}^{(l,1)}_{t};\mathbf{SA}^{(l,2)}_{t};...;\mathbf{SA}^{(l,H)}_{t}]\mathbf{W}^{(l)}\in\mathbb{R}^{T^\prime\times D}.
\end{equation}
The output $\mathbf{MSA}^{(l)}_{t}$ is then merged with the input embedding via the skip connection. The 2023 baseline processes the merged signal by the dropout layer and layer normalization (LN) as
\begin{equation}
\mathbf{e}^{(l+1)}_{t}=\mathrm{LN} \left[\mathrm{Dropout} \left[ \mathbf{MSA}^{(l)}_t+\mathbf{e}^{(l)}_{t}\right]\right] \in\mathbb{R}^{T^\prime\times D}.
\end{equation} 

\subsubsection{Feed forward structure}
In the architecture of the ResNet-Conformer and EINV2 models, which utilize the Conformer framework as a decoder, LN is applied before the MHSA mechanism, while dropout is employed subsequently. The input to the LN is subsequently integrated with the features resulting from the dropout operation. Moreover, the Conformer decoder incorporates a convolutional module positioned after the MHSA mechanism, which facilitates the fusion of channel information through the application of pointwise and one-dimensional depthwise convolutional layers. The Conformer decoder further includes two half-dimensional feed-forward (FFW) modules that envelop the MHSA and convolutional components. Notwithstanding the inclusion of convolutional and FFW modules within the Conformer decoder, the architecture remains consistent with the 2023 baseline in its application of the MHSA exclusively to temporal sequences. Then, the output temporal sequences pass the fully-connected (FC) layers to match the multi-ACCDOA output.

\subsection{Loss function}
All three previous models utilize the multi-ACCDOA output format and an auxiliary duplicating permutation invariant training (ADPIT) in order to resolve the track permutation between estimation and the ground truth outputs. Considering all possible track permutations as $Perm$, one possible frame-level permutation at $t$ and class $c$ is $\alpha\in Perm[ct]$. Then the ADPIT loss using multi-ACCDOA becomes
\begin{equation} \label{Eq:multiACCDOAPIT}
\mathcal{L}^{ADPIT}=\frac{1}{CT}\sum^C_{c=1}\sum^T_{t=1} \min\limits_{\alpha\in Perm[ct]}{l}^{ACCDOA}_{\alpha,ct}
\end{equation} where $T$ is the total number of time bins of the output, $C$ is the total number of classes and ${l}^{ACCDOA}_{\alpha,ct}$ is the mean-squared error (MSE) loss of ACCDOA in permutation $\alpha$ and frame $t$ and class $c$ which is defined as 
\begin{equation} \label{Eq:multiACCDOA_loss}
{l}^{ACCDOA}_{\alpha,ct}=\frac{1}{N}\sum^N_{n=1}\lVert \mathbf{P}_{nct}-\mathbf{\hat{P}}_{\alpha,nct} \rVert^2_2,
\end{equation} where $N$ is the total number of tracks, $\mathbf{P}_{nct}$ is the target multi-ACCDOA label of track $n$, class $c$ and time $t$ and $\mathbf{\hat{P}}_{nct}$ is the corresponding estimated multi-ACCDOA. The lowest loss for each frame is back-propagated for training.

\subsection{Inference}
For inference, the similarity of tracks is considered to determine the number of sound sources among $N$ tracks of multi-ACCDOA output~\cite{multiaccdoa}. First, the similarity of estimated coordinates between different tracks is calculated for every frame $t$. If none of the tracks of class $c$ have cosine similarity within 15$\degree$, there are $N$ estimated sources because the tracks that do not have similarity within 15$\degree$ with other tracks are considered different sound sources. However, if any two tracks are within 15$\degree$, for class $c$ in time $t$, they are considered as same sound source. If all tracks of class $c$ at time $t$ share similarities with each other, then it is treated as a single sound source. Then, the estimated sound sources with vector lengths greater than 0.5 are taken as estimated DoAs. 


\section{CST-former: Transformer with Channel-Spectro-Temporal Attention}
\label{sec:proposed}
\begin{figure*}
\includegraphics[width=\linewidth]{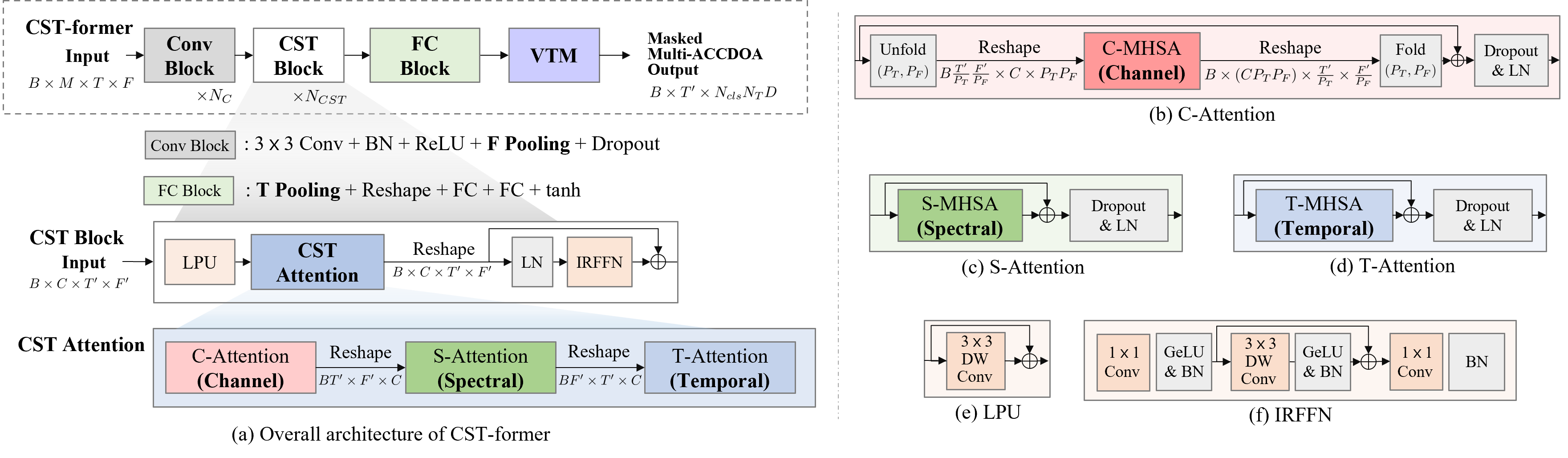}
\caption{Overview and details of CST-former architecture with channel attention layer utilizing unfolded local embedding.}
\label{fig:Architecture}
\vspace*{-3mm}
\end{figure*}

The overall architecture of the proposed model is presented in Fig.~\ref{fig:Architecture}. The model consists of ConvBlock encoders encoding input data, CST blocks extracting features through multidimensional attention, a fully-connected (FC) Block mapping features into multi-ACCDOA data, and VTM adaptively masking FN estimations.  

\subsection{Encoder}
The input data of Eq.~\ref{Eq:input} is encoded through three cascaded ConvBlock encoders, resulting in $C$-channel features $\mathbf{z}
\in{\mathbb{R}^{C\times T^\prime\! \times F^\prime}}$ 
with $T^\prime$ time frames and $F^\prime$ frequency bins.
Each ConvBlock comprises $3\times3$ convolution layers followed by batch normalization (BN), rectified linear unit (ReLU) activation, time-frequency (T-F) pooling, and dropout layers. The $T^\prime$ and $F^\prime$ are determined with the kernels of max pooling as described in Table~\ref{tab:TFPooling}. 

\begin{table}
\caption{Time-frequency (T-F) pooling configurations of baseline models and CST-former}
\label{tab:TFPooling}
\vspace*{-3mm}
\begin{center}
\resizebox{0.9\columnwidth}{!}{
\begin{tabular}{c|c|c|c|c} 
\hline
Model & Baseline~\cite{2022DCASE_baseline, dcase2023_baseline}& Front & Middle & End \\ \hline \hline
ConvBlock Layer 1 & (5,4) & (5,2) & (1,1) & (1,1)\\ \hline
ConvBlock Layer 2 & (1,2) & (1,2) & (1,2) & (1,2)\\ \hline
ConvBlock Layer 3 & (1,2) & (1,1) & (5,2) & (1,2)\\ \hline
FC Block (T-pooling) & - & - & -& (5,1)\\ \hline
\end{tabular}
}\vspace*{-5mm}
\end{center}
\end{table}

\begin{table}
\caption{
Kernel configurations for ULE on each CST Block layer\\ when $N_{CST}=2$ or $N_{CST}=4$}
\label{tab:MSULE}
\vspace*{-3mm}
\begin{center}
\begin{tabular}{c|c|c|c|c} 
\hline
 $N_{CST}$ & 1 & 2 & 3 & 4 \\ \hline \hline
\multirow{2}{*}{ULE ($P_T$, $P_F$)}& (10,4)& (10,4)& -&-\\ \cline{2-5}
& (25,4) & (25,4) & (25,4) & (25,4) \\ \hline
MSULE ($P_T$, $P_F$)& (25,4) & (10,4) & (5,4) & (5,2)\\ \hline
\end{tabular}
\vspace*{-5mm}
\end{center}
\end{table}

\subsection{Decoder: CST Block}
The encoded features are then processed by a decoder which is a series of CST Blocks, repeated $N_{CST}$ times. Each CST Block contains CST Attention sandwiched by the local perception unit (LPU) and inverted residual feed-forward network (IRFFN). The LPU and IRFFN processing the input and output of the CST Attention are adapted from those used for Convolution Meets Transformer (CMT) architecture~\cite{CMT_2022_CVPR} for its benefits in extracting and aggregating local T-F features.

The LPU extracts local time and frequency features using a $3\times3$ depth-wise convolution (DWConv) layer. Then, the input of LPU is added with the residual connection. Therefore, the output of LPU layer in $l$-th layer, $\mathbf{z}^{(l)}_\mathrm{LPU}$ 
becomes
\begin{equation} \label{Eq:LPU}
\mathbf{z}^{(l)}
_{\mathrm{LPU}}=\mathrm{DWConv_{(3,3)}}\left[ \mathbf{z}
\right] + \mathbf{z}
\in \mathbb{R}^{C\times T^\prime \times F^\prime}
\end{equation} where the $\mathrm{DWConv}$ has kernel size of $(3,3)$. 

The IRFFN consists of an expansion layer with a four-fold increase in dimensions, a depth-wise convolution with a residual connection, and a projection layer.
The IRFFN replaces the linear FFN of conformer~\cite{Conformer} with convolutional expansion and projection layers, allowing the model to aggregate local T-F features using the $3\times3$ DWConv layer.

To apply IRFFN after the CST Attention, the output $\mathbf{e}^{\prime(l)}_{t}$ of $l$-th CST Attention layer is reshaped to
\begin{equation} \label{Eq:IRFFN_input}
\mathbf{z}^{(l)}
_{\mathrm{IRFFN}}=\mathrm{Reshape}\left[ \mathbf{e}^{\prime(l)}_{t} \right] \in \mathbb{R}^{C\times T^\prime \times F^\prime},
\end{equation} which passes the IRFFN to learn the local relation and shuffle the channel information.

After LN, the channel dimension is expanded four times with $1\times 1$ convolution (Conv) followed by Gaussian error linear unit (GeLU) activation function and BN as
\begin{equation} \label{Eq:IRFFN_channel_expansion}
\mathbf{z}^{\prime(l)}
_\mathrm{IRFFN}
=\mathrm{BN}\left[\mathrm{GeLU}\left[\mathrm{Conv_{(1,1)}}\left[\mathrm{LN}\left[\mathbf{z}^{(l)}
_{\mathrm{IRFFN}}\right]\right]\right]\right],
\end{equation}
where $\mathbf{z}^{\prime(l)}_{\mathrm{IRFFN}}\in \mathbb{R}^{4C\times T^\prime \times F^\prime}$.
This channel-expanded feature passes a $3\times 3$ DWConv layer followed by GeLU, BN, and residual connection as
\begin{equation} \label{Eq:IRFFN_depthwise}
\mathbf{z}^{\prime\prime(l)}_{\mathrm{IRFFN}}=\mathrm{BN}\left[\mathrm{GeLU}\left[\mathrm{DWConv_{(3,3)}}\left[\mathbf{z}^{\prime(l)}_{\mathrm{IRFFN}}\right]\right]\right]+\mathbf{z}^{\prime(l)}_{\mathrm{IRFFN}},
\end{equation} 
where $\mathbf{z}^{\prime\prime(l)}_{\mathrm{IRFFN}} \in \mathbb{R}^{4C\times T^\prime \times F^\prime}$, 
which enables the model to learn the local T-F relation.
Then, the channel dimension is reconstructed to that of CST Attention with the second $1\times1$ convolution layer followed by another BN and residual connection.
\begin{equation} \label{Eq:IRFFN_depthwise}
\mathbf{z}^{\prime\prime\prime(l)}_{\mathrm{IRFFN}}=\mathrm{BN}\left[\mathrm{Conv_{(1,1)}}\left[\mathbf{z}^{\prime\prime(l)}\right]\right]+\mathbf{z}^{(l)}_{\mathrm{IRFFN}},
\end{equation} where $\mathbf{z}^{\prime\prime\prime(l)}_{\mathrm{IRFFN}}\in \mathbb{R}^{C\times T^\prime \times F^\prime}$.
By expanding the channel dimension and reducing it to the original size, the channel information is shuffled in a suitable order for the next layer. In addition, the local T-F relation can be trained with $3\times3$ DWConv while the CST Attention layer focuses on the global contexts of each three domains.

\subsection{Channel Attention with Unfolded Local Embedding}
The CST Attention layer in the CST Block incorporates three distinct attention mechanisms, each tailored to address a specific domain: channel (C-Attention), spectral (S-Attention), and temporal (T-Attention). Previous work on DST attention~\cite{ST_attention} employed separate spectral and temporal attention layers, utilizing the encoded channel dimension as the embedding dimension for both spectral and temporal attention. However, implementing channel attention (CA) across the channel dimension necessitates an additional embedding dimension. Therefore, we propose a method for generating embeddings for CA. 

Instead of directly using the spatial information in the channel dimension as the sequence for C-Attention, a novel embedding generation method, unfolded local embedding (ULE), is suggested for CST-former. 
To achieve this, we convert the non-overlapped local T-F patches of size $(P_T,\!P_F)$ into embeddings using the ULE method.

Then, the encoded channel dimension $C$ is used as the channel sequence for CA. ULE utilizes local temporal and spectral information obtained by unfolding the local T-F bins in the kernel of size $(P_T,P_F)$ and stacking the unfolded local information in channel dimension as 
\begin{equation} \label{Eq:unfold}
\mathbf{e}^{unfold}_{ch} = \mathrm{Unfold}_{(P_T,P_F)}\left[ \mathbf{z}^{(l)}_{
_\mathrm{LPU}} \right] \in \mathbb{R}^{CP_T P_F\times \frac{T^\prime}{P_T} \times \frac{F^\prime}{P_F}}.
\end{equation} where the $\frac{T^\prime}{P_T}$ and $\frac{F^\prime}{P_F}$ are the remaining global context of the temporal and spectral domain, respectively.

For CA, the unfolded feature $\mathbf{e}^{unfold}_{ch}$ is reshaped by allocating local T-F bins of size $D_{local}=(P_T P_F)$ to the embedding dimension and the remaining global context to the batch dimension, while using the channel information as a sequence. This results in the batch dimension of $(B\frac{T^\prime}{P_T}\frac{F^\prime}{P_F})$ and the embedding feature given by 
\begin{equation} \label{Eq:uleinput}
\mathbf{e}^{(l)}_{ch} = \mathrm{Reshape}\left[ \mathbf{e}^{unfold}_{ch} \right] \in \mathbb{R}^{C \times D_{local}}.
\end{equation} Using $\mathbf{e}^{(l)}_{ch}$ for CA, the size of the CA map becomes $C\times C$ .

\begin{figure}
  \centering
    \includegraphics[width=\linewidth]{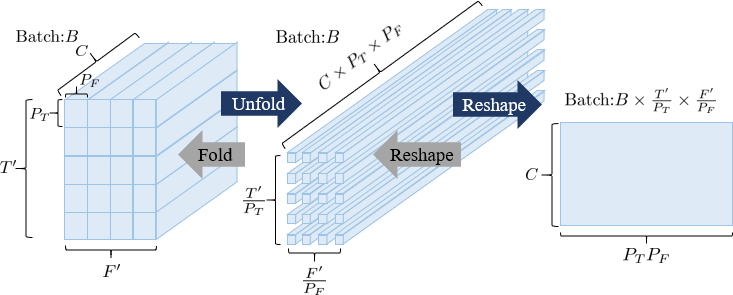}
    \caption{Generation and restoration of unfolded local embedding for channel attention using unfold and fold layer.} 
    \label{fig:Unfold-Fold}
\end{figure}

After being processed by CA, the output is reshaped to its original size by reshaping and folding the features, which is the inverse of the unfolding process, to get input embedding for spectral attention as shown in Fig.~\ref{fig:Unfold-Fold}. Therefore, the reshaped feature can be written as 
\begin{equation}
\mathbf{e}^{\prime(l)}_{ch,re} = \mathrm{Fold}_{(P_T,P_F)}\left[\mathrm{Reshape}\left[\mathbf{e}^{\prime (l)}_{ch}\right]\right] \in \mathbb{R}^{C\times T^\prime \times F^\prime}.
\end{equation}
The reshaped output is combined with the input of the attention layer using a residual connection, followed by dropout and LN.

The size of the ULE kernel ($P_{T}$, $P_{F}$) in CST Block determines the T-F resolution of embeddings and computation cost. 
For example, using a smaller ULE kernel reduces the embedding dimension for channel attention and yields more global T-F bins $\left(\frac{T^\prime}{P_T},\frac{F^\prime}{P_F}\right)$, inducing finer resolution in T-F domain. Finer T-F resolution of embeddings may solve temporal inconsistency problems during inference but may hamper capturing the global context in embeddings.
Therefore, varying kernel sizes along the cascaded CST Blocks can help the model learn both global and local localization features. This technique, denoted as multiscale ULE (MSULE), is tested on the proposed CST-former with subsequently reduced kernels (Table~\ref{tab:MSULE}). 

\subsection{Spectral and Temporal Attention}
For the subsequent spectral (S-Attention) and temporal (T-Attention) attentions, the input dimensions are permuted to $(F^\prime,C)$ and $(T^\prime,C)$, respectively, with the remaining temporal and spectral dimensions allocated to the batch dimension. As presented in Fig.~\ref{fig:Architecture}(c) and (d), MHSA with residual connection is then applied, followed by dropout and LN.
The input and output of the CST Attention are processed through LPU and IRFFN, respectively.

\subsection{Output Extractor}
In the FC Block, the data is reshaped to $(T^\prime,F^\prime C)$ and projected to multi-ACCDOA vectors by an optional time pooling layer, two subsequent FC layers, and $\mathrm{tanh}$ activation function. The obtained multi-ACCDOA vectors $\mathbf{\hat{P}}_{nct}=\alpha_{nct}\mathbf{d}_{nct}\in \mathbb{R}^D$ are the estimation of the ground truth multi-ACCDOA vectors ($\mathbf{P}_{nct}$) in a $D$-dimensional space for the track number $n\in \{1,...,N_T\}$, class index $c\in\{1,...,N_{cls}\}$, and time index $t\in \{1,...,T^\prime\}$. Here, $\alpha_{nct}\in [0,1]$ is the length of a vector indicating the occurrence probability of a sound event, and $\mathbf{d}_{nct}\in \mathbb{R}^D$ is the unit vector representing DoA.

\begin{figure}
  \centering
    \includegraphics[width=\linewidth]{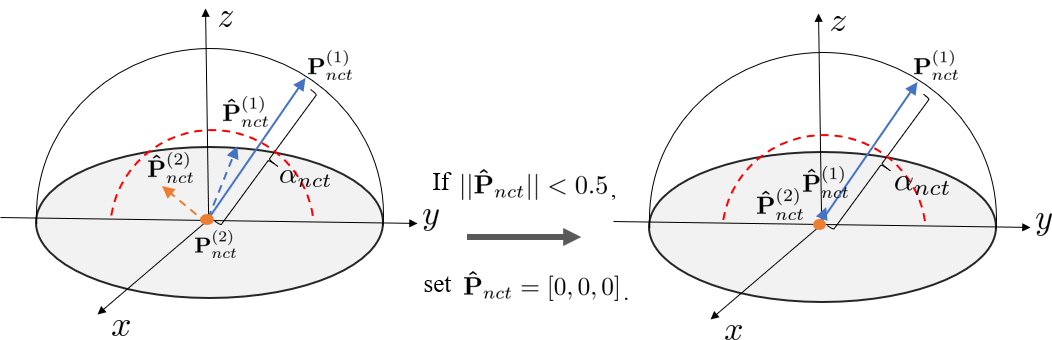}
    \caption{Visualization of vector threshold masking. $\mathbf{P}_{nct}$ is the ground truth multi-ACCDOA of track $n$ and class $c$ in time $t$, and $\mathbf{\hat{P}}_{nct}$ is the estimated vector output. The red dashed line is the SED threshold. $\mathbf{\hat{P}}_{nct}$ is masked when the vector length is shorter than the SED threshold.} 
    \label{fig:VTM}
\end{figure}

\subsection{Finetuning with Vector Threshold Masking (VTM)}

At the inference time, the occurrence of sound events is determined by comparing the length of a multi-ACCDOA vector $\alpha_{nct}$ with a fixed threshold of $0.5$. However, the performance of event detection is evaluated by the MSE of multi-ACCDOA during training. This discrepancy inevitably produces FNs when the estimated multi-ACCDOA vector of an existing sound event has a length less than $0.5$ ($\alpha_{nct}<0.5$). 

To reduce FNs, a training-time masking technique called vector threshold masking (VTM) is applied before calculating the MSE loss (Fig.~\ref{fig:VTM}). 
VTM masks the output when the estimated vector of any individual track, class, and time is below the threshold $0.5$. This forces the model to amplify the vector length of potential FNs (e.g. $\mathbf{\hat{P}}^{(1)}_{nct}$ in Fig.~\ref{fig:VTM}) by enlarging the MSE loss. 
With VTM, the MSE loss of Eq.~\ref{Eq:multiACCDOA_loss} is modified to
\begin{equation}\label{Eq:VTM}
{l}^{VTM}_{\alpha,t}=\frac{1}{NC}\sum^N_{n=1}\sum^C_{c=1}\lVert\mathbf{{P}}_{nct}-\mathrm{VTM}(\mathbf{\hat{P}}_{nct})\rVert^2_2,
\end{equation}
where $\mathrm{VTM}(\mathbf{\hat{P}}_{nct})$ is the masked output of the estimated multi-ACCDOA ($\mathbf{\hat{P}}_{nct}$) and $\mathbf{{P}}_{nct}$ is the ground truth of track $n$, class $c$, and time frame $t$.
To avoid training inefficiency, the VTM is implemented by finetuning the model pre-trained without VTM using the loss of Eq.~\ref{Eq:VTM}.

\subsection{Inference-time Processing (IP)}
\subsubsection{Inference Overlapping (IO)}
Training over a limited dataset often induces jitters or inconsistencies in the estimated SELD parameters. A representative example is the inconsistency across the relative location of a time frame within input sequences. To address this issue, the IO technique estimating the SELD parameters of a single time frame multiple times from multiple sequences including the time frame of interest at different locations was proposed in ~\cite{IO}. These sequences were prepared by sectioning long input data using windows with a hop size of $1\,$s and a sequence length $5\,$s. Five different estimations resulting from five different sequences can be obtained, and the median value of the estimated multi-ACCDOAs is taken as the output.

\begin{figure*}
  \centering
    \includegraphics[width=0.9\textwidth]{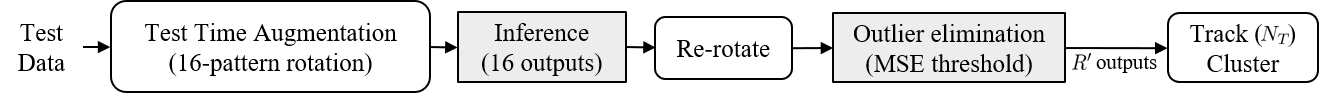}
    \caption{Process of clustered-track augmented inference (CTAI).} 
    \label{fig:CTAI}
\end{figure*}

\begin{figure}
  \centering
    \includegraphics[width=0.8\linewidth]{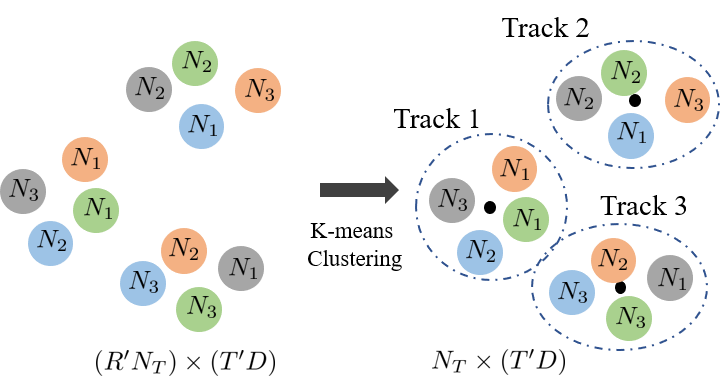}
    \caption{An example of clustering $R^\prime=4$ inference outputs of rotated input to $N_T=3$ tracks using K-means clustering.} 
    \label{fig:CTAI_2}
\end{figure}

\subsubsection{Clustered-track Augmented Inference (CTAI)}
Another example of inconsistency is the rotational inconsistencies. To secure consistency over coordinate rotations, audio channel swapping (ACS) augmentation~\cite{ACS_TASLP} has been popularly utilized. ACS swaps channels and signs of FoA signals to generate $R$ rotation-augmented data. Re-rotating inferred DoAs of rotation-augmented data back to the original coordinates should exhibit consistent estimation of DoAs. ACS has also been employed for the test time augmentation (TTA)~\cite{dcase2022_3rd}, which clusters multiple estimations from ACS at inference time and chooses the cluster center (average) as the final estimation. 

CTAI proposed in this work utilizes a similar principle but clusters both tracks and rotations to mitigate possible track permutation caused by TTA. 
To eliminate the outliers among inference outputs of augmented data, the MSEs are calculated between re-rotated outputs of $R$ rotations and the output from the original input. $R^\prime$ estimations with MSEs less than a threshold ($10^{-3}$) are used for clustering.

Then, 
flattened multi-ACCDOA vectors of a size $T^\prime D$, obtained from $N_T$ tracks of $R^\prime$ estimations, are clustered as illustrated in Fig.~\ref{fig:CTAI_2}.
For individual classes $N_{cls}$, the K-means clustering is applied independently 
to cluster $N_T$ track clusters among the $R^\prime N_T$ sequences. 
The initial cluster center was randomly chosen and the maximum iteration was 500.
Then, the cluster centers of the clustered tracks are regarded as the estimated tracks of time and DoA, $T^\prime D$. Then, the CTAI output is reshaped to the multi-ACCDOA output dimension $(T^\prime,N_{cls}N_TD)$ for evaluation .

\section{Experiments and Result Analysis} \label{sec:results}
\subsection{Implementation Details}\label{ssec:implementation}
For training, \emph{dev-set-train} of STARSS22 or STARSS23 was utilized, as well as \emph{synth-set}. For evaluation, \emph{dev-set-test} of them was used. 
We trained the proposed model utilizing the FoA format signals. We applied the short-time Fourier transform (STFT) with a hop size of $0.02$\,s and a window size of $0.04$\,s on the audio signals sampled at $24$\,kHz. The spectrograms were transformed into log-mel scales using 64 mel filter banks. Also, IVs were calculated by multiplying the complex conjugation of the omnidirectional spectrograms with the other directional array spectrograms and taking the real part of the multiplication. These IVs were normalized using the total energy of the signals. The total number of input channels $M$ was seven consisting of four original spectrograms and three intensity components. 

For comparison with various models, multi-ACCDOA was used as the output format, and the number of tracks $N_T$ was $3$. The input signals were segmented into $5$\,s without overlapping for training. The number of Conv Blocks $N_C$ was three, and the number of filters for 2D convolution $C$ was $64$ or $128$. The number of CST Blocks $N_{CST}$ was two (Small or Base), four (Large), and six (Huge). All MHSAs had eight multi-heads. 
$P_T$ and $P_F$ were set as presented in Table~\ref{tab:MSULE} in terms of uniscale and multiscale. 

The model with $C=64$ was trained for $500$ epochs while $200$ epochs for $C=128$, with mean-squared error (MSE) loss and a batch size of $32$. The Adam optimizer was employed. Moreover, a tri-stage learning rate scheduler which comprises a ramp and a cosine scheduler was used with an upper limit of $10^{-3}$ for models with $C=64$ and $10^{-4}$ for $C=128$.

We evaluated our method using the SELD score ($\mathcal{S}_{SELD}$)~\cite{2018DCASE_baseline}, which is an average of four metrics: location-dependent error rate (ER) and F1-score (F1) with $20^\circ$ angular threshold, as well as class-dependent localization error (LE) and localization recall (LR). The F1, LE, and LR were macro-averaged across individual classes.

For training with a limited dataset, we employed various augmentations, including frameshift~\cite{freqshift} shifting the input feature and label along the time axis using a randomly determined shifting parameter. Additionally, time masking~\cite{timemasking} was implemented by randomly masking features and DoA labels along the time axis. Furthermore, ACS was utilized for spatial augmentation. Lastly, moderate mix-up~\cite{M-mixup} was incorporated to blend two input features within a batch, which is an enhanced version of mix-up~\cite{mixup} specifically designed for the SELD task, randomly selecting a mixing ratio from beta distribution to blend two features and taking the label of larger ratio as the target. 
All four augmentation methods were randomly selected and applied for every batch and iteration.

\begin{table}
\caption{Parameter studies on CST-former utilizing STARSS23. Bold and underlined numbers indicate the best and second-best models, respectively. Size parameters are described in the form of (T-F pooling type, $N_{cst}$, ~$C$).} 
\label{tab:param_study}
\vspace*{-3mm}
\begin{center}
\resizebox{1\columnwidth}{!}{
\setlength\tabcolsep{0.2em}
\begin{tabular}[width=\textwidth]{p{1.1cm}|p{2.3cm}|c|cccc}
\hline 
Param. & Variations&$\mathcal{S}_{SELD}\downarrow$&$\mathrm{ER}_{20^\circ}\downarrow$&$\mathrm{F1}_{20^\circ}\uparrow$&$\mathrm{LE}_{CD}\downarrow$&$\mathrm{LR}_{CD}\uparrow$ \\\hline \hline
\multirow{4}{*}{Size} & Small\,(Front,\,2,\,64) & 0.3738 & 0.46 & 47.3 & 16.2 & 58.2 \\
& Base\,(Middle,\,2,\,64)& 0.3594& \textbf{0.42} & 49.2 & \underline{15.4} & 57.6  \\ 
 & \textbf{Large}\,(End,\,4,\,128)&\textbf{0.3323} & 
 \underline{0.43}& \underline{52.3}& \textbf{14.8} & \textbf{66.0} \\ 
& Huge\,(End,\,6,\,128)& \underline{0.3447} & 0.45 & \textbf{52.2} & 15.6 &  \underline{63.9}  \\ \hline
\multirow{4}{*}{Attention} & CS-former& 0.4529 & 0.64& 36.4&21.9 &  58.6\\ 
& CT-former& \underline{0.3449} & \textbf{0.43} & \underline{50.0}& 16.1 & \underline{64.0} \\ 
 & ST-former & 0.3556 & 0.47& \underline{50.0}& \underline{15.2} & 63.2  \\ 
 & \textbf{CST-former} & \textbf{0.3323} & \textbf{0.43}& \textbf{52.3}& \textbf{14.8} & \textbf{66.0} \\ \hline
 \multirow{3}{*}{Order} & TCS & \underline{0.3407}& {0.44} & 51.0 & \underline{14.9}& \underline{65.0}\\ 
 & SCT & 0.3408 & {0.44}& \underline{52.0} & 15.0 & 64.0  \\ 
 & \textbf{CST} & \textbf{0.3323}& \textbf{0.43}& \textbf{52.3}&\textbf{14.8} & \textbf{66.0} \\ \hline
 \multirow{2}{*}{MSULE}& ULE& 0.3323 & \textbf{0.43}& 52.3& 14.8 &{66.0} \\
 & \textbf{MSULE}& \textbf{0.3317}& 0.44& \textbf{53.1}& \textbf{14.7}&\textbf{66.4}\\\hline
\end{tabular}
}
\vspace*{-5mm}
\end{center}
\end{table}

\subsection{Parameter Studies of CST-former}
We conducted in-depth parameter studies for CST-former, and results are presented in Table~\ref{tab:param_study}. The models were trained with {\em dev-set-train} of STARSS23 and {\em synth-set}, and evaluated on {\em dev-set-test} of STARSS23.

\subsubsection{Model size (Size)}
Models of varying sizes were evaluated using different T-F pooling configurations in Table~\ref{tab:TFPooling}, the number of CST Blocks ($N_{CST}$), and embedding dimensions ($C$). Among these, the CST-former (Large) model, configured with $N_{CST}=4$, $C=128$, and T-F pooling at End, demonstrated the best performance. This outcome suggests that performing T-F pooling in the later layers enhances feature resolution in the convolutional layer, enabling the model to capture more information and improving SELD performance. Furthermore, the results from the CST-former (Large) model indicate that increasing model size generally benefits SELD tasks. However, excessively large models, such as CST-former (Huge), occasionally exhibit performance degradation due to overfitting to the training data.


\subsubsection{Ablation of Attentions (Attention)}
To evaluate the effectiveness of the proposed multidimensional attention mechanism, each of the three attention components was ablated individually. All ablations resulted in a decline in overall performance metrics, with the most significant degradation observed when T-Attention was removed. This underscores the critical role of T-Attention in SELD performance, particularly as the target output must preserve temporal order. The second most substantial decline was with the ablation of C-Attention. 

Notably, the removal of S- or C-Attention did not independently impact SED or localization metrics. However, the performance benefits of S- and C-Attention are interconnected, influencing both SED and localization outcomes. 
Nevertheless, the inclusion of all three domain-specific attention mechanisms provided the most substantial improvement, achieving a $\mathcal{S}_{SELD}$ of $0.3323$ with better ER, F1, LE, and LR metrics than any attention ablations. This highlights the significance of multidimensional attention in the SELD task, as it facilitates the extraction of valuable domain information from a limited dataset. 


\begin{table}
\caption{Ablation studies of VTM finetuning, IP methods, and inference sequence length on CST-former (Large) with MSULE conducted over STARSS23. Bold and underlined numbers indicate the best and second-best models, respectively.} 
\label{tab:ablation_2}
\vspace*{-4mm}
\begin{center}
\resizebox{1\columnwidth}{!}{
\setlength\tabcolsep{0.2em}
\begin{tabular}[width=\textwidth]{c|c|c|c|c|cccc}
\hline 
\multirow{2}{*}{VTM} & \multicolumn{2}{c|}{IP}& \multirow{2}{*}{SL (s)} &\multirow{2}{*}{$\mathcal{S}_{SELD}\downarrow$}&\multirow{2}{*}{$\mathrm{ER}_{20^\circ}\downarrow$}&\multirow{2}{*}{$\mathrm{F1}_{20^\circ}\uparrow$}&\multirow{2}{*}{$\mathrm{LE}_{CD}\downarrow$}&\multirow{2}{*}{$\mathrm{LR}_{CD}\uparrow$} \\ \cline{2-3}
& IO & CTAI & & & & & &  \\ \hline \hline
&& & \multirow{5}{*}{5}& 0.3317& 0.44& 53.1& 14.7& 66.4 \\ 
\ding{52}&& &&0.3299&0.45&53.0&14.7&68.2 \\ 
\ding{52}&\ding{52}&& & {0.3161}& \underline{0.42}& {56.4}& {13.9}& 66.9 \\ 
\ding{52}&&\ding{52}& & 0.3549& 0.58& 50.0& 14.5& \textbf{74.1}\\ 
\ding{52}&\ding{52}&\ding{52}& & \underline{0.3111}& \underline{0.42}& \underline{56.8}& \textbf{13.6}& \underline{68.3}\\ \hline
\ding{52}&\ding{52}& \ding{52}& \textbf{10}& \textbf{0.3067}& \textbf{0.41}& \textbf{57.7}& \underline{13.8}&\underline{68.3}\\
\ding{52}&\ding{52}& \ding{52}& 20& 0.3224& 0.41& 55.6& 14.0&64.2
\\ \hline
\end{tabular}
}
\vspace*{-5mm}
\end{center}
\end{table}

\subsubsection{Order of Attention (Order)}

The sequence of T-, S-, and C-Attention layers was varied to identify the optimal ordering. While the order had a relatively minor impact on overall SELD performance, it influenced specific metrics, particularly F1 and LR. The optimal sequence was determined to be CST. This arrangement is advantageous because the local time-frequency (T-F) embeddings may undergo reorganization within the C-Attention layer. Consequently, placing spectral and temporal attention (S- and T-Attention) after C-Attention helps preserve the contextual information, facilitating more accurate estimations.

\subsubsection{Multiscale ULE}
The parameter studies of multiscale ULE on STARSS23 is presented in Table.~\ref{tab:param_study} (MSULE). It is notable that both of the localization accuracy, $\mathrm{LE}_{CD}$ and $\mathrm{LR}_{CD}$, and the localization-dependent F1 ($\mathrm{F1}_{20^\circ}$) performance improve with multiscale ULE. Therefore, the $\mathcal{S}_{SELD}$ improves from $0.3323$ to $0.3317$ by applying multiscale ULE to CST-former (Large). This indicates the diminishing T-F kernels of ULE increases the T-F resolution and affects the SELD performance in a positive way. 



\begin{table*}
\caption{Performance comparison of various SELD models on STARSS22. 
\\Bold and underlined numbers indicate the best and second-best models, respectively.} 
\label{tab:comparison_sota_2022}
\centering  
\resizebox{1.65\columnwidth}{!}{
\setlength\tabcolsep{0.5em}
\begin{tabular}[width=\textwidth]{l|c|c|cccc}
\hline 
 Model & Params &$\mathcal{S}_{SELD}\downarrow$&$\mathrm{ER}_{20^\circ}\downarrow$&$\mathrm{F1}_{20^\circ}\uparrow$&$\mathrm{LE}_{CD}\downarrow$&$\mathrm{LR}_{CD}\uparrow$ \\\hline \hline
 2023 baseline~\cite{dcase2023_baseline} & 0.74M&0.4146&0.58&43.1&20.2&60.3\\ 
EINV2~\cite{einv2_2022} &85.3M & 0.4073 (1.7$\%\downarrow$)& 0.56 & 42.4 & 19.3  & 61.4 \\ 
MFF-EINV2~\cite{mff_einv2} & 26.9M & {0.3910} (5.7$\%\downarrow$)&0.58&48.0&18.7&64.0\\
CST-former (Base) &0.39M & 0.3773 (9.0$\%\downarrow$)& 0.54& 46.8 & 18.9 & 66.8\\ 
ResNet-Conformer\footnotemark~\cite{ResnetConformer}& 13.1M& 0.3591 (13.4$\%\downarrow$)& 0.51& 51.0 & {17.4} & 66.0 \\ \hline\textbf{CST-former (Large)}& \multirow{3}{*}{1.88M} & 0.3440 (17.0$\%\downarrow$)& {0.49}& 52.1& 18.2 & 69.4\\ 
+\textbf{VTM} &  & {0.3373} (18.6$\%\downarrow$)& {0.49}& {53.6}& 17.7 & \underline{70.3}\\
+\textbf{VTM}+\textbf{IP}  &  & {0.3222} (22.3$\%\downarrow$)& \underline{0.45}& {56.6}& {17.0} &{69.0}\\  \cline{2-7}
+\textbf{MSULE} & \multirow{4}{*}{1.77M} & {0.3398} (18.0$\%\downarrow$)& {0.49}& {52.6}& 16.1 & {69.2}\\
+\textbf{MSULE}+\textbf{VTM} &  & {0.3347} (19.3$\%\downarrow$)& {0.49}& {53.9}& {16.0}& {70.1}\\
+\textbf{MSULE}+\textbf{VTM}+\textbf{IP}  (SL 5 s)  &  & \underline{0.3148} (24.1$\%\downarrow$)& {0.46}& \underline{57.3}& \underline{15.7}&\textbf{71.5}\\
+\textbf{MSULE}+\textbf{VTM}+\textbf{IP}  (SL 10 s)&  & \textbf{0.3066} (26.0$\%\downarrow$)& \textbf{0.42}& \textbf{59.7}& \textbf{15.6}&{68.4}\\
\hline
\multicolumn{7}{l}{$^{1}$Only ACS is applied for data augmentation while all other models apply identical augmentation methods.}
\end{tabular}
}
\end{table*}

\subsection{Finetuning and Inference-time Processing}
Ablation studies on the additional techniques to enhance the proposed CST-former model, VTM and IP, are presented in Table~\ref{tab:ablation_2}. Supposing the benefits of MSULE on CST-former compared to ULE, the effects of using VTM and IP containing IO and CTAI are demonstrated on CST-former (Large) with MSULE utilizing STARSS23. Additionally, an ablation study examining the impact of sequence length during the inference phase is provided.

\subsubsection{Use of Vector Threshold Masking (VTM)}
Finetuning the pre-trained CST-former with VTM improves LR while maintaining LE but deteriorating the SED performance slightly in both ER and F1. This implies that VTM improves localization accuracy, by decreasing the number of FNs thereby increasing true positives.

\subsubsection{Inference-time Processing (IP)}
When IP techniques were applied to the model finetuned with VTM, IO improved SED (ER and F1) performance and angular estimation accuracy while degrading LR. Meanwhile, CTAI is beneficial to angular accuracy especially on LR. Therefore, combining IO and CTAI improves both SED and DoAE performances compared to without IP, lowering the $\mathcal{S}_{SELD}$ from $0.3299$ to $0.3111$.

\subsubsection{Inference Sequence Length (SL)}
The input sequence length for inference affects the model performance. With longer sequence length, more global temporal context can be evaluated within T-Attention of CST-former. By enlarging the sequence length from $5$\,s to $10$\,s, the performance improves from $0.3111$ to $0.3067$. However, the longer sequence does not always guarantee better performance as the performance slightly deteriorates with the length of $20$\,s.

\begin{table*}
\caption{Performance comparison of various SELD models on STARSS23 with augmentation and additional data for training.\\ Bold and underlined numbers indicate the best and second-best models, respectively.} 
\label{tab:comparison_sota_2023}
\centering  
\resizebox{1.9\columnwidth}{!}{
\setlength\tabcolsep{0.5em}
\begin{tabular}{c|c|c|c|c|cccc}
\hline 
Team & Model & \begin{tabular}{c} Additional\\Data\end{tabular} & \begin{tabular}{c} Post\\Processing\end{tabular} & $\mathcal{S}_{SELD}\downarrow$&$\mathrm{ER}_{20^\circ}\downarrow$&$\mathrm{F1}_{20^\circ}\uparrow$&$\mathrm{LE}_{CD}\downarrow$&$\mathrm{LR}_{CD}\uparrow$ \\ \hline \hline
0&2023 baseline~\cite{dcase2023_baseline} &-&- &0.4138&0.53&41.3&17.7&56.0\\
1&ResNet-Conformer~\cite{dcase2023_1st}&\ding{52}&\ding{52}&\textbf{0.2992} ($27.6\%\downarrow$)& 0.44 & \textbf{58.0} & \textbf{13.8} & \textbf{74.0}\\ 
&Proposed Model &-&\ding{52} & \underline{0.3067} ($25.9\%\downarrow$)& \textbf{0.41}& \underline{57.7}& \textbf{13.8}&\underline{68.3}\\  
2&ResNet-Conformer~\cite{dcase2023_2nd} &\ding{52}& \ding{52} & 0.3209 ($22.5\%\downarrow$)& 0.43& 54.8 & 14.7 & 68.0 \\ 
3& ResNet-Conformer \cite{dcase2023_4th} & \ding{52} & -  & 0.3255 ($21.3\%\downarrow$)& \underline{0.42} & 55.4 & 15.0  & 64.8 \\ 
4& EINV2 \cite{dcase2023_3rd} & \ding{52} & -  & 0.3633 ($12.2\%\downarrow$)& 0.48 & 48.0 & 16.2  & 63.7 \\ 
5& Conv-Conformer \cite{dcase2023_6th} & \ding{52}  &-  & 0.3677 ($11.1\%\downarrow$)& 0.50 & 49.1 & 16.5  & 63.0 \\  \hline
\end{tabular}
}
\end{table*}

\subsection{Comparison to Existing SELD Models}
Our preliminary study~\cite{CST_former} demonstrates the effectiveness of CST-former compared to the other models when trained without augmentation. The performance comparison of various SELD models trained with data augmentation is required and, therefore, presented in this research.
The comparisons of the proposed model to various existing SELD models~\cite{ResnetConformer, 2022DCASE_baseline,einv2_2022, mff_einv2} 
are presented in Table~\ref{tab:comparison_sota_2022} and~\ref{tab:comparison_sota_2023}.
\subsubsection{STARSS22}
Table~\ref{tab:comparison_sota_2022} shows the results of models that were trained with {\em dev-set-train} of STARSS22 and {\em synth-set}, and evaluated on {\em dev-set-test} of STARSS22. 
Additional ablation studies on MSULE, VTM, and IP are performed to verify the impacts of proposed methods on STARSS22.

CST-former (Large) presents the best SELD performance, outperforming EINV2 and ResNet-Conformer even without VTM, achieving $\mathcal{S}_{SELD}$ of $0.3440$. The performance is further enhanced with VTM and IP, and $\mathcal{S}_{SELD}$ is lowered to $0.3373$ with only VTM and to $0.3222$ with additional IP, achieving the best scores on all four metrics compared to previous SOTA models, presenting the effectiveness of the proposed methods 
on both SED and DoAE tasks.

Also, MSULE on CST-former (Large) further enhances the SELD performance on STARSS22. All the cases of implementing VTM and IP on CST-former (Large) with MSULE increase SELD performance, decreasing the $\mathcal{S}_{SELD}$ to $0.3148$ with improved F1, LE, and LR compared to the result without MSULE where  $\mathcal{S}_{SELD}$ is $0.3222$. Also, the change of inference SL from $5$\,s to $10$\,s enhances performances with most of the metrics except for the LR, diminishing $\mathcal{S}_{SELD}$ to $0.3066$. 

Moreover, it can be inferred from Table~\ref{tab:comparison_sota_2022}, that the proposed model has a fairly small number of parameters compared to the other models while achieving the best SELD performance. The total number of parameters from CST-former (Large) decreases with MSULE even when the performance increases. 
This proves that the suggested architecture with the proposed multidimensional attention-based transformer extracts the dimensional information efficiently from the given datasets even with a limited number of data.

\subsubsection{STARSS23}
The performance comparison on STARSS23 with additional synthetic datasets is presented in Table~\ref{tab:comparison_sota_2023}. The results are from the submission of the 2023 DCASE challenge task 3. For a fair comparison, the results are presented utilizing FoA format signals transformed to log-mel spectrograms and IVs, and without a model ensemble. All the other models except for the CST-former and 2023 baseline were trained with additional datasets generated individually. 

The performance of the proposed model even without any additional data is $0.3067$ achieving 25.9\% of improvements compared to the 2023 baseline, which is still fairly comparable to the performance of other models trained with additional datasets, achieving the second-best performance with respect to the $\mathcal{S}_{SELD}$. Especially, the proposed model shows strengths in SED with the best ER and second-best F1, and in localization with the best LE and second-best LR. 

\begin{table*}[hbt]
\caption{Duration of each class data of STARSS23.} 
\label{tab:classtypes}
\begin{center}
\resizebox{2\columnwidth}{!}{
\setlength\tabcolsep{0.2em}
\begin{tabular}{c|c|c|c|c|c|c|c|c|c|c|c|c|c|c} 
\hline 
\multicolumn{2}{c|}{}&  1& 2& 3& 4& 5& 6& 7& 8& 9&10& 11 &12 &13\\ \hline
\multicolumn{2}{c|}{Class Type}&\begin{tabular}{c} Female\\Speech\end{tabular} & \begin{tabular}{c} Male\\Speech\end{tabular} & Clapping & Telephone& Laughter &  \begin{tabular}{c}Domestic\\Sounds\end{tabular}& Footsteps & Door & Music& \begin{tabular}{c}Musical \\Instrument\end{tabular}&\begin{tabular}{c}Water\\ Tap\end{tabular}&Bell & Knock \\ \hline
\multirow{2}{*}{\begin{tabular}{c} Duration\\(min)\end{tabular}}& Train &74.0&  82.8& 1.4& 2.5& 8.9& 47.6& 5.5& 1.7& 66.4& 7.4&1.4&2.5 & 0.2\\ \cline{2-15}
& Test & 45.2&  66.4& 1.2& 1.5& 6.7& 27.2& 7.7& 0.9& 52.5& 43.1&4.5&2.0&0.1\\ \hline
\end{tabular}
}
\end{center}
\end{table*}

\begin{figure}
\includegraphics[width=\columnwidth]{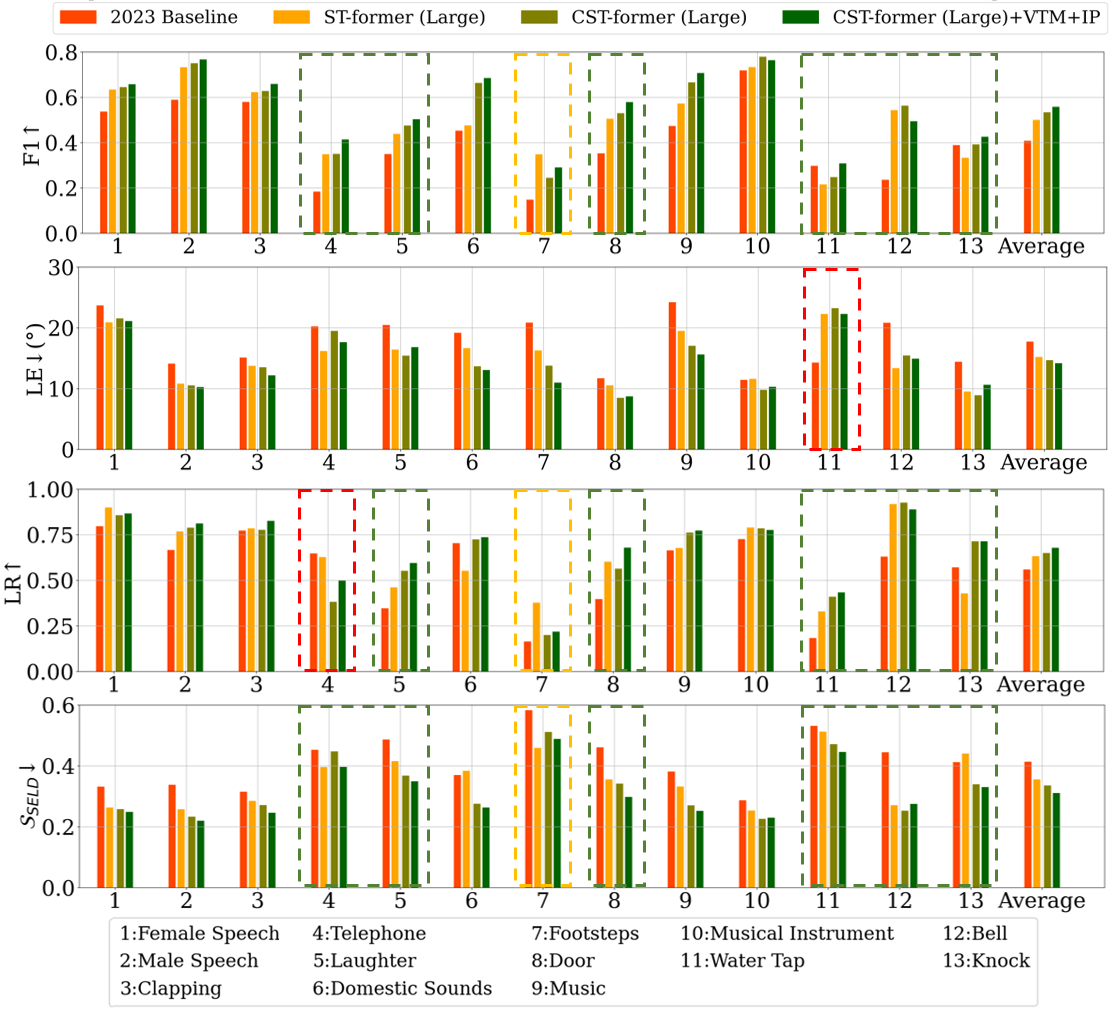}
\caption{Per class results of F1$\uparrow$ (first row), LE$\downarrow$ (second row), LR$\uparrow$ (third row), and $\mathcal{S}_{SELD}\downarrow$ (fourth row) on {\em dev-set-test} of STARSS23.} 
\label{fig:classwise}
\end{figure}

\begin{figure}
\includegraphics[width=\columnwidth]{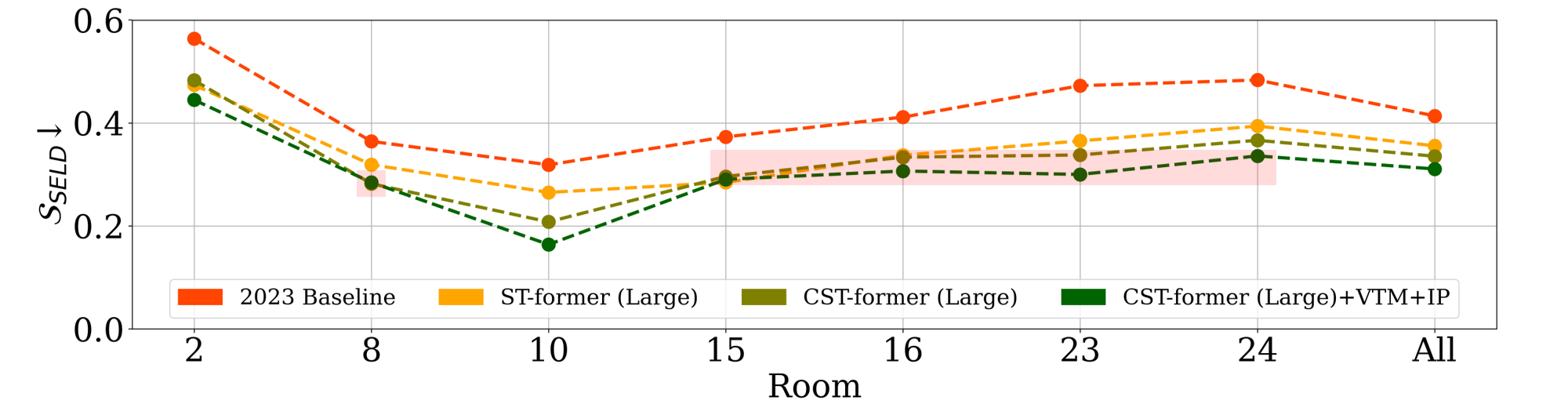}
\vspace*{-6mm}
\caption{Per room $\mathcal{S}_{SELD}\downarrow$ on {\em dev-set-test} of STARSS23.} 
\label{fig:roomwise}
\end{figure}

\section{Discussion} \label{sec:discussion}
\subsection{Class and Room Dependencies}
There is an imbalance in the number of real-recorded data across different classes as detailed in Table~\ref{tab:classtypes}. Certain types of data with impulsive characteristics 
are rare in numbers compared to long-duration signals. 
This class imbalance causes the overfitting of models to dominant classes and deteriorates the SELD performance especially when metrics are macro-averaged over classes.
Moreover, the models are tested on datasets recorded at rooms that are unseen during training, and these room variations diversify the RIRs, SNR, and reverberation time, differing the characteristics of the test data from those of the training data. 

Therefore, the SELD performance of the proposed model is analyzed in-depth across different classes and room environments in Fig. \ref{fig:classwise} and Fig. \ref{fig:roomwise}, presenting the effectiveness of the proposed multidimensional attention, VTM, and IP. The comparison sets are the 2023 baseline with temporal RNN and attention layers (temporal), ST-former (spectral and temporal), and the proposed CST-former (spatial, spectral, and temporal) with and without VTM and IP. All models were trained and tested by the STARSS23, and with the same augmentation methods as described in section \ref{ssec:implementation}.

\subsubsection{Class Dependency}
Class-specific performance variation due to class imbalance and impulsive characteristics exists. However, the per class F1, LE, LR, and $\mathcal{S}_{SELD}$ presented in Fig.~\ref{fig:classwise} demonstrate the tendency of performance enhancement with multidimensional attention. 

Classes with impulsive signals (4, 5, 7, 8, 11\,--\,13, except 3) present relatively bad F1 or LR compared to other classes. The best-performing model with each metric is presented with a dashed box for each class. Most of the classes with impulsiveness show improved F1 and LR with more multidimensional attention. 
The number of classes with F1 over 0.6 increases with multidimensional attention along with VTM and IP. 
The LE is also improved with the proposed methods; eight classes presenting LE under $15^\circ$. Also, seven classes have an LR of over 0.8 with CST-former. These affect the localization-dependent SED metrics positively. 

Notably, the proposed method achieves remarkable $\mathcal{S}_{SELD}$ improvement in classes 5, 8, 11\,--\,13. 
It can be inferred that more dimensions in the attention layer and the proposed VTM finetuning help improve the performances on the classes with impulsive signals that are short in duration, therefore, mitigating the per-class variations.

\begin{figure*}
\centering
\includegraphics[width=2\columnwidth]{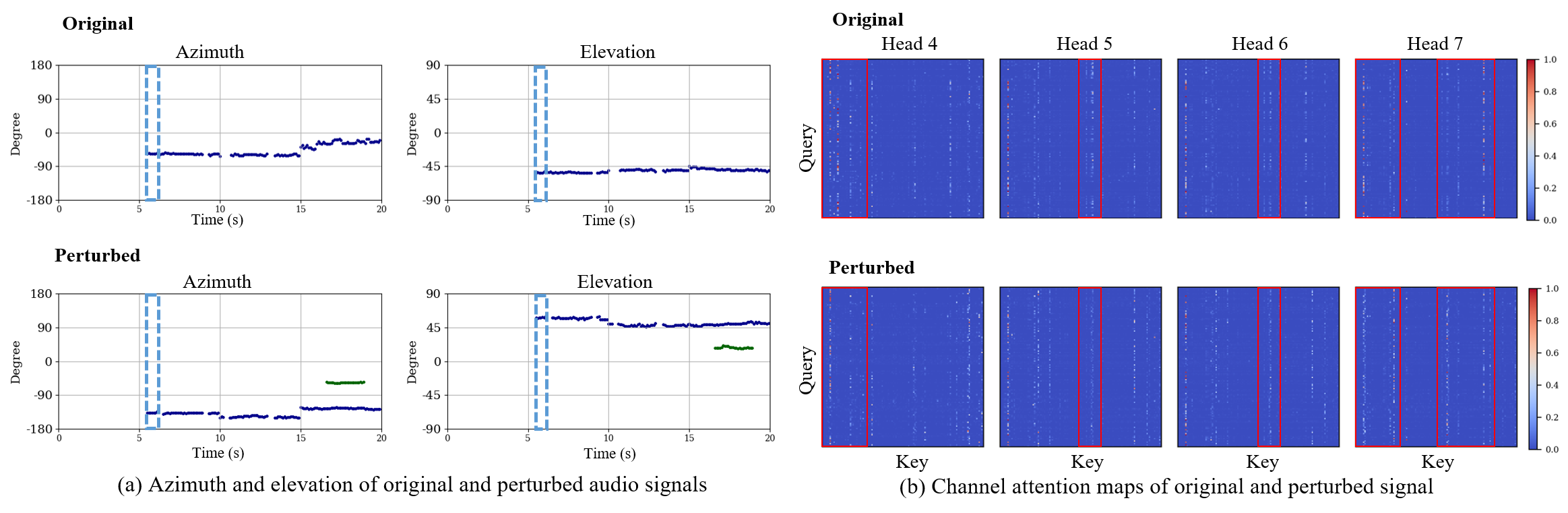}
\caption{Channel attention with source location perturbation. For source location perturbation, azimuth is decreased by $-90^\circ$, and elevation is reversed. Channel attention maps of the last CST Block layer on time frames of the blue box in (a) are depicted in (b) for the original and perturbed sources. Changes in channel attention are highlighted with red boxes.}
\label{fig:Ch_location_perturbation}
\end{figure*}

\subsubsection{Room Dependency}
The per room $\mathcal{S}_{SELD}$ presented in Fig.~\ref{fig:roomwise} denotes the advantage of the proposed method in various room environments. Because the room environments of {\em dev-set-test} differ from {\em dev-set-train}, the model overfitted to the train dataset can perform badly in various unseen environments of {\em dev-set-test}. For example, the 2023 baseline in Fig.~\ref{fig:roomwise} presents varying and low performances in different unseen rooms. The deterioration due to diverse environments is mitigated with ST-former and further alleviated with CST-former, and with VTM and IP, lowering the $\mathcal{S}_{SELD}$ to a similar level in rooms 8, 15, 16, 23, and 24. However, the degradation in room 2 still requires more improvement in future work.

\subsection{Attention Map Analysis}

\subsubsection{Source Location Perturbation}
To verify the role of CA, the channel attention maps with sound source perturbation are compared in Fig.~\ref{fig:Ch_location_perturbation}. For source location perturbation, azimuth is decreased by $-90^\circ$, and elevation is reversed. CA maps of the last CST Block layer in time frames marked in the blue dashed box in Fig.~\ref{fig:Ch_location_perturbation}(a) are depicted in Fig.~\ref{fig:Ch_location_perturbation}(b). Because DoAs are the only different factor in this comparison, the changes in CA maps of original and perturbed signals presented in Fig.~\ref{fig:Ch_location_perturbation}(b) are solely due to the location differences. Notably, the locations of vertical lines in CA maps are different and the highlighted pattern of vertical lines is changed. Therefore, the location perturbation affects the channel attention and the channel attention aggregates the localization-related features. 

\subsubsection{Cosine Similarity}
\begin{figure}
  \centering
    \includegraphics[width=1\columnwidth]{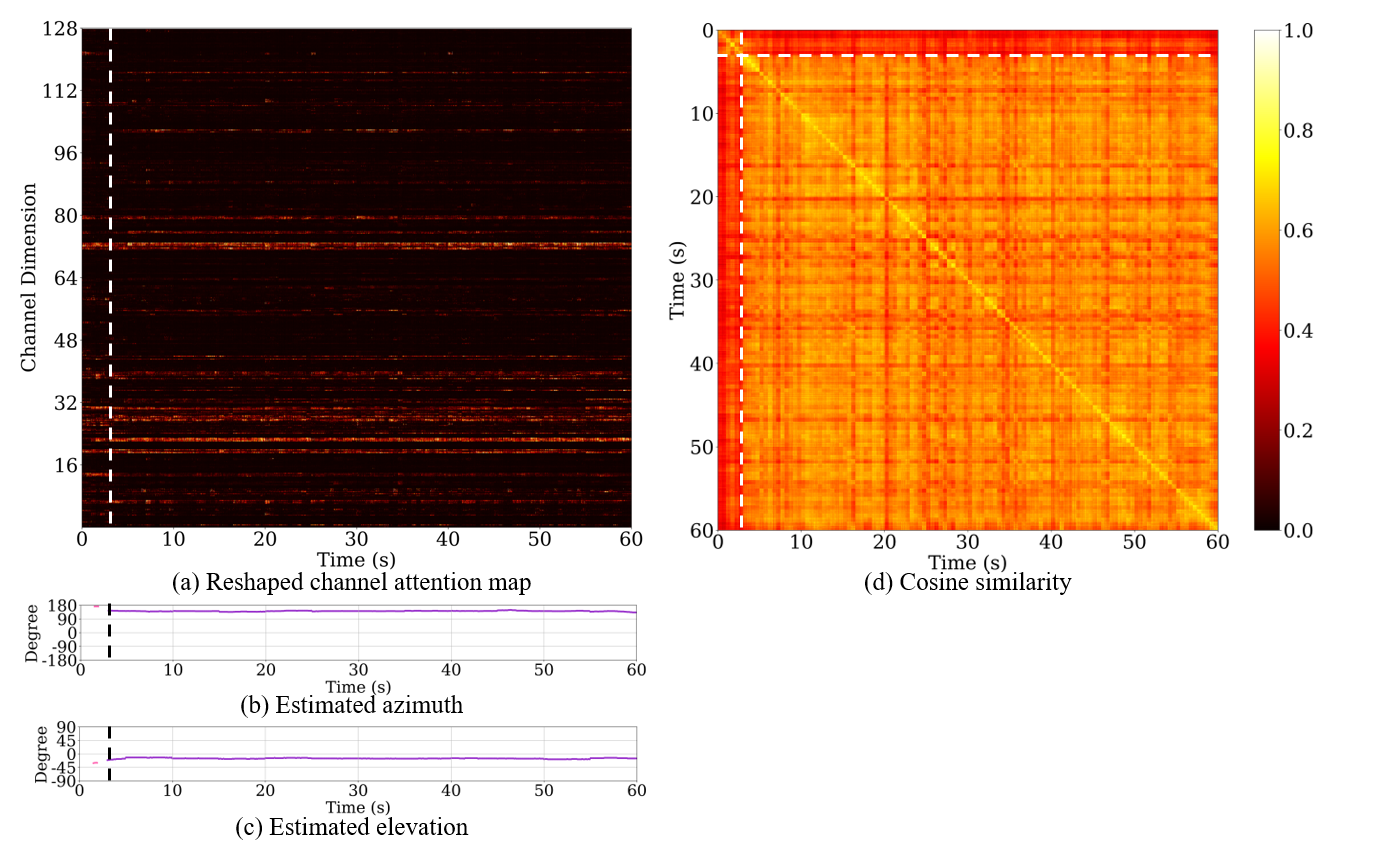}
    \caption{An example of a stationary source 
    (a) Reshaped channel attention map (b) the predicted azimuth (c) the estimated elevation of a stationary source signal, and (d) the cosine similarity of the reshaped channel attention map in the temporal domain.}
    \label{fig:ch_attn_1_stationary}
\end{figure}

\begin{figure}
  \centering
    \includegraphics[width=1\columnwidth]{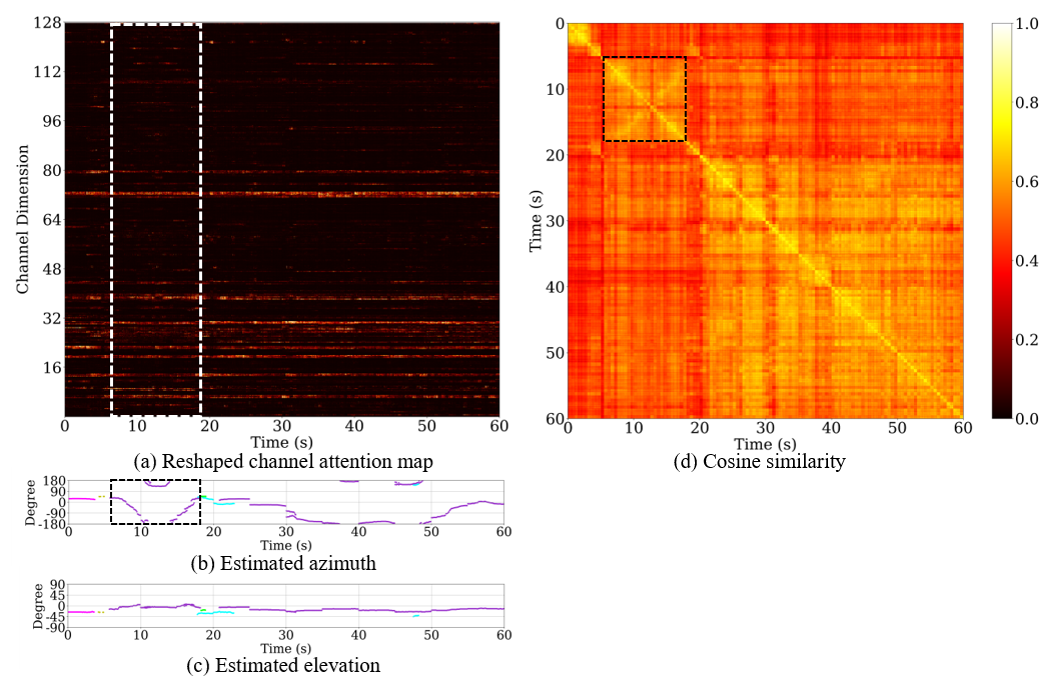}
    \caption{An example of a transient sound source that constantly changes location over time. (a) Reshaped channel attention map, (b) the predicted azimuth, (c) the estimated elevation of multiple moving source signals, and (d) the cosine similarity of the reshaped channel attention map in the temporal domain.} 
    \label{fig:ch_attn_2_moving}
\end{figure}

To further demonstrate the effects of CA on localization, the cosine similarities of channel attention maps in different time frames are calculated and visualized in Fig.~\ref{fig:ch_attn_1_stationary} and Fig.~\ref{fig:ch_attn_2_moving}.

Taking the maximum values among multiple heads $h$, the remaining dimension of CA maps is $[(B\frac{T^\prime}{P_T} \frac{F^\prime}{P_F}), C_q,C_k]$, where $C_q$ and $C_k$ are equal to the size of $C$. Vertical features are distinguished in different $C_k$ bins as shown in Fig. ~\ref{fig:Ch_location_perturbation}(b). 
To visualize the CA characteristics due to temporal features, the remaining CA 
is reshaped to $[(C_k \frac{F^\prime}{P_F}), (B \frac{T^\prime}{P_T} C_q)]$ so that the $C_q$ can be displayed over the local time bins and the verticality of the global frequency-wise channel dimension and the spectral features of $(C_k \frac{F^\prime}{P_F})$ can be depicted horizontally on the y-axis. The batch $B$ during inference contains the segments divided into input sequence lengths from an audio signal, inducing $(B \frac{T^\prime}{P_T} C_q)$ to have temporal information. In Fig.~\ref{fig:ch_attn_1_stationary} and Fig.~\ref{fig:ch_attn_2_moving}, examples of CA maps reshaped to $[(C_k \frac{F^\prime}{P_F}), (B \frac{T^\prime}{P_T} C_q)]$ are demonstrated. For each case, the cosine similarity of the reshaped CA vector $(C_k \frac{F^\prime}{P_F})$ over $(B \frac{T^\prime}{P_T} C_q)$ is presented to verify that the source location is related to the channel attention.

The first example in Fig.~\ref{fig:ch_attn_1_stationary} is of a stationary sound source, where a musical instrument (purple) is played over time in a fixed location. The horizontally displayed spectrum of the channel attention map also presents stationary characteristics. Only a slight temporal difference appears when the audio starts around 3\,s where the dashed line is located. In Fig.~\ref{fig:ch_attn_1_stationary}(d), the cosine similarities between the reshaped channel attention map of different time frames are presented. The time frames with stationary sound share similar characteristics on the channel attention map presenting high cosine similarities. Meanwhile, a low similarity is captured between the frames without any sound source and the frames with the stationary source, as depicted in Fig.~\ref{fig:ch_attn_1_stationary}(d). It can be induced that the CA remains highly similar between time frames if the class and location of the sound source remain the same over time.

On the contrary, the channel attention map in Fig.~\ref{fig:ch_attn_2_moving} presents a transient spectrum as the sound sources move continuously in the azimuthal direction. Compared to the stationary source, the highlights of the reshaped channel attention map in the dashed box change along with the time axis. This phenomenon is also demonstrated in the black dashed box of the cosine similarity in Fig.~\ref{fig:ch_attn_2_moving}(d). From 6\,s to 18\,s, the azimuthal direction is changed sinusoidally and the corresponding cosine similarity presents a shape of X, showing that frames with similar azimuth at 6\,s and 18\,s have high similarity and the similarity decreases as the azimuth changes. This intuitively indicates the role of CA as an aid on localization. 

Moreover, the frames in 0\,--\,4\,s that are different in class (pink) from the other time frames (purple or cyan) have very low cosine similarities with the frames after 4\,s. This also implies that the class information is reflected in the CA and global frequency bins. Therefore, the CA of the proposed CST-former with ULE affects both the classification and localization.

\section{Conclusion} \label{sec:conclusion}
A transformer with separate multidimensional attention on the channel-spectro-temporal domains was suggested for the SELD task in real scenes. The ULE was validated as a critical component for channel attention, utilizing the unfolded local temporal and spectral embedding as the embedding. 
Analysis of the attention maps confirms the advantages of multidimensional attention for the SELD task, with channel attention aiding source localization, demonstrating its effectiveness. 
Moreover, the newly proposed techniques such as MSULE, VTM, and IP with IO and CTAI were proven to be beneficial for the SELD task.

The CST-former proposes a possibility of overcoming the challenges due to the difficulties of real data acquisition faced by SELD in real scenes, by achieving comparable results to those of the top-tier models trained with additionally generated data, establishing its effectiveness as a DNN architecture tailored for SELD in real scenes. The proposed methods achieve substantially high performances on STARSS22 and STARSS23 even without any external datasets for training, mitigating performance degradation due to class imbalance and various room environments. The effects of the proposed multidimensional attention on the SELD are noteworthy.


\end{document}